\documentclass{article}

\usepackage{PRIMEarxiv}

\usepackage[utf8]{inputenc} 
\usepackage[T1]{fontenc}    
\usepackage{hyperref}       
\usepackage{url}            
\usepackage{booktabs}       
\usepackage{amsfonts}       
\usepackage{nicefrac}       
\usepackage{microtype}      
\usepackage{lipsum}
\usepackage{fancyhdr}       
\usepackage{graphicx}       
\graphicspath{{media/}}     
\usepackage{amsmath} 
\usepackage{multirow}
\usepackage{url}
\usepackage{bbding} 
\usepackage{makecell}
\usepackage{graphicx}
\usepackage{float}
\title{Autoregressive Enzyme Function Prediction with Multi-scale Multi-modality Fusion
\thanks{Preprint. Work in progress.} 
}
\author{
  Dingyi Rong, Wenzhuo Zheng, Bozitao Zhong, Zhouhan Lin, Liang Hong, Ning Liu \\
  Shanghai Jiaotong University \\
  \texttt{\{r892546826, darkcorvus, hongl3liang, ningliu\}@sjtu.edu.cn} \\
  \texttt{\{zbztzhz, lin.zhouhan\}@gmail.com}
}

\begin{document}
\maketitle

\begin{abstract}
Accurate prediction of enzyme function is crucial for elucidating biological mechanisms and driving innovation across various sectors. Existing deep learning methods tend to rely solely on either sequence data or structural data and predict the EC number as a whole, neglecting the intrinsic hierarchical structure of EC numbers. To address these limitations, we introduce MAPred, a novel multi-modality and multi-scale model designed to autoregressively predict the EC number of proteins. MAPred integrates both the primary amino acid sequence and the 3D tokens of proteins, employing a dual-pathway approach to capture comprehensive protein characteristics and essential local functional sites. Additionally, MAPred utilizes an autoregressive prediction network to sequentially predict the digits of the EC number, leveraging the hierarchical organization of EC classifications. Evaluations on benchmark datasets, including New-392, Price, and New-815, demonstrate that our method outperforms existing models, marking a significant advance in the reliability and granularity of protein function prediction within bioinformatics.
\end{abstract}


\section{Introduction}

In the realm of bioinformatics, accurately determining the functions of enzymes utilizing Enzyme Commission (EC) numbers \cite{bairoch2000enzyme} is a long-standing and challenging task. This process can offer insights into their catalytic mechanisms, substrate specificities, and potential applications in various industries \cite{nagels2012production, manning2010stability, dupuis2023precision, nyyssola2022role}. However, the ability to experimentally determine the EC numbers of enzymes is significantly lagging behind the rapid pace at which enzyme data is being generated \cite{zhou2019cafa}, for the experimental determination of function is a complex, time-consuming, and resource-intensive process that involves multiple steps \cite{price2018mutant}. Therefore, developing computational methods for predicting EC numbers is particularly important.

The traditional bioinformatics methods, including sequence alignment-based approaches like BLASTp \cite{altschul1990basic}, have been crucial in predicting protein EC numbers. However, they heavily rely on pre-existing knowledge stored in databases \cite{li2019blastp, liu2011cuda, mcginnis2004blast}, which limits their capacity to accurately predict novel proteins lacking close homologs. Moreover, the complexity of protein evolution challenges the correlation between sequence similarity and functional conservation, leading to inaccuracies in predictions.

In recent years, deep learning-based approaches have emerged as a promising alternative in the field of EC number prediction, offering potential solutions to the limitations of traditional methods, yet they still face their own limitations. On one hand, most existing approaches rely solely on either protein sequence \cite{deepec,ProteInfer,UDSMProt, DeepECtransformer} or structure information \cite{amidi2018enzynet,derevyanko2018deep,zhang2023enhancing,gligorijevic2021structure} for predictions, facing their respective limitations: sequence-only models lack the critical three-dimensional context necessary for function, potentially misinterpreting proteins with similar sequences but different structures, while structure-only models struggle due to the challenges in obtaining protein structures and the inability to capture dynamic conformational changes necessary for activity. 
On the other hand, current approaches treat the EC numbers as a singular entity, relying on the protein's overall characteristics for their prediction \cite{li2018deepre,liu2017deep,amidi2018enzynet,hermosilla2020intrinsic}. While focusing on global features is valuable for capturing the overall information of a protein, it may overlook critical local regions within the protein. Moreover, treating EC numbers as a singular entity overlooks their inherent hierarchical organization. An enzyme's EC number is a concatenation of four numerical codes, each indicating different levels of specificity in substrate and reaction type \cite{bairoch2000enzyme}. Ignoring this hierarchy can lead to a loss of information.

In this work, we propose MAPred (\textbf{M}ulti-scale multi-modality \textbf{A}utoregressive \textbf{P}redictor) to overcome the above problems, which features three designs. 
First, to obtain a joint representation of the protein's primary and tertiary structures, given that 3Di alphabet \cite{foldseek} can discretize the structure into a set of sequences with strong positional correlation to the protein sequence and can serve as a simplified substitute for protein structure, we derive the corresponding 3Di tokens from the protein sequence using ProstT5 \cite{heinzinger2023prostt5}, which serves as a combined input to our network. 
Second, to capture both the global characteristics of the protein and the local functional site features, we have developed a hybrid feature extraction network that includes both a global feature extraction pathway and a local feature extraction pathway. 
Considering that the input is composed of two parts, it is logical for the global feature extraction block to employ an interlaced sequence-3Di cross-attention mechanism, and in the local feature extraction block, CNN is well-suited for extracting local structural information. 
Third, recognizing the strong sequential dependencies among the four digits of the EC number, inspired by \cite{wehrmann2018hierarchical}, MAPred utilizes an autoregressive prediction network to predict each of the four digits sequentially. 
As far as we know, MAPred is the first to use a hybrid feature of protein sequences and 3Di to predict the EC number, and we conduct extensive ablation studies to reveal the role of each component, which helps deepen the reader’s understanding of MAPred and may provide further inspiration for subsequent research. In summary, our contributions include: 

\begin{itemize}
\item We combined the protein sequence and its 3Di tokens as inputs to the model, incorporating both the primary sequence and tertiary structure information.
\item We propose a novel hybrid feature extraction network to learn global and local representations from multi-modality protein representations.
\item We propose an autoregressive label prediction network, which establishes a sequential prediction logic by creating a hierarchy for Enzyme Commission (EC) number predictions.
\item We comprehensively compare advanced models in real-world datasets, e.g., New-392, Price, and New-815, and demonstrate the potential of the components we proposed.
\end{itemize}

\section{Related Work}
\paragraph{Sequence similarity-based methods} 

Sequence similarity-based methods, particularly those utilizing sequence alignment tools \cite{BLAST, PSI-BLAST}, rely on the comparison of a query protein sequence against a database of sequences with known functions to find matches with high sequence similarity. The effectiveness of these methods hinges on the quality of the sequence alignment and annotation quality of the databases used \cite{boutet2007uniprotkb, schomburg2002brenda}. Recent developments have integrated traditional homology searches with more sophisticated algorithms. For instance, methods employing Hidden Markov Models (HMMs) \cite{eddy1996hidden}, as implemented in HMMER \cite{finn2011hmmer} and Gapped-BLAST \cite{PSI-BLAST}, use probabilistic models of sequence evolution to detect homologous protein families and predict their function, including EC numbers, even in cases of low sequence similarity.

\paragraph{Sequence-based ML methods} 
Sequence-based ML methods employ neural networks to capture the inherent patterns within amino acid sequences to predict the EC numbers of enzymes. The earliest attempts used Convolutional Neural Networks (CNNs) to capture local sequence motifs that are indicative of enzymatic function \cite{deepec,ProteInfer,UDSMProt}. Following the success of CNNs, Recurrent Neural Networks (RNNs), particularly Long Short-Term Memory (LSTM) networks, were explored due to their ability to capture long-range dependencies in sequences \cite{li2018deepre,liu2017deep,elhaj2021deep_cnn_lstm_go}. This capability is crucial for understanding the full context of enzyme functions, as determinants of enzymatic activity can be dispersed throughout the protein sequence. Recently, Transformer-based methods marked a significant milestone in protein function prediction. These methods \cite{EnzBert,DeepECtransformer,yu2023enzyme,brandes2022proteinbert,rives2021biological} leverage self-attention mechanisms to capture complex relationships between amino acids, regardless of their distance in the sequence, and achieve state-of-the-art performance on benchmark datasets.

\paragraph{Structure-based ML methods} 
To delve deeper into the interactions between amino acids at each position within proteins, some methods utilize the protein's structure as input for the model. As the tertiary structure of proteins reveals, amino acids that are distant in the sequence can be closely positioned in the spatial arrangement. Convolutional neural networks (CNNs) have shown promising results in capturing the intricate patterns within protein structures. These approaches typically represent the input structure as Grids \cite{amidi2018enzynet,derevyanko2018deep} or Residue Contact Map \cite{gao2019prediction}, and utilize multiple convolutional layers to capture hierarchical features and learned representations that capture crucial structural information relevant to enzymatic function. Recent innovations have also explored the use of graph neural networks (GNNs) for this task. By representing protein structures as graphs, where nodes represent amino acids and edges represent their interactions, GNN-based models \cite{hermosilla2020intrinsic,hermosilla2022contrastive,zhang2023enhancing,zhang2023protein,gligorijevic2021structure} have been particularly adept at capturing the relational information within proteins.

\section{Methodology}
\label{headings}

\subsection{Overall Framework}
\begin{figure}[h]
    \centering
    \includegraphics[width=0.95\linewidth]{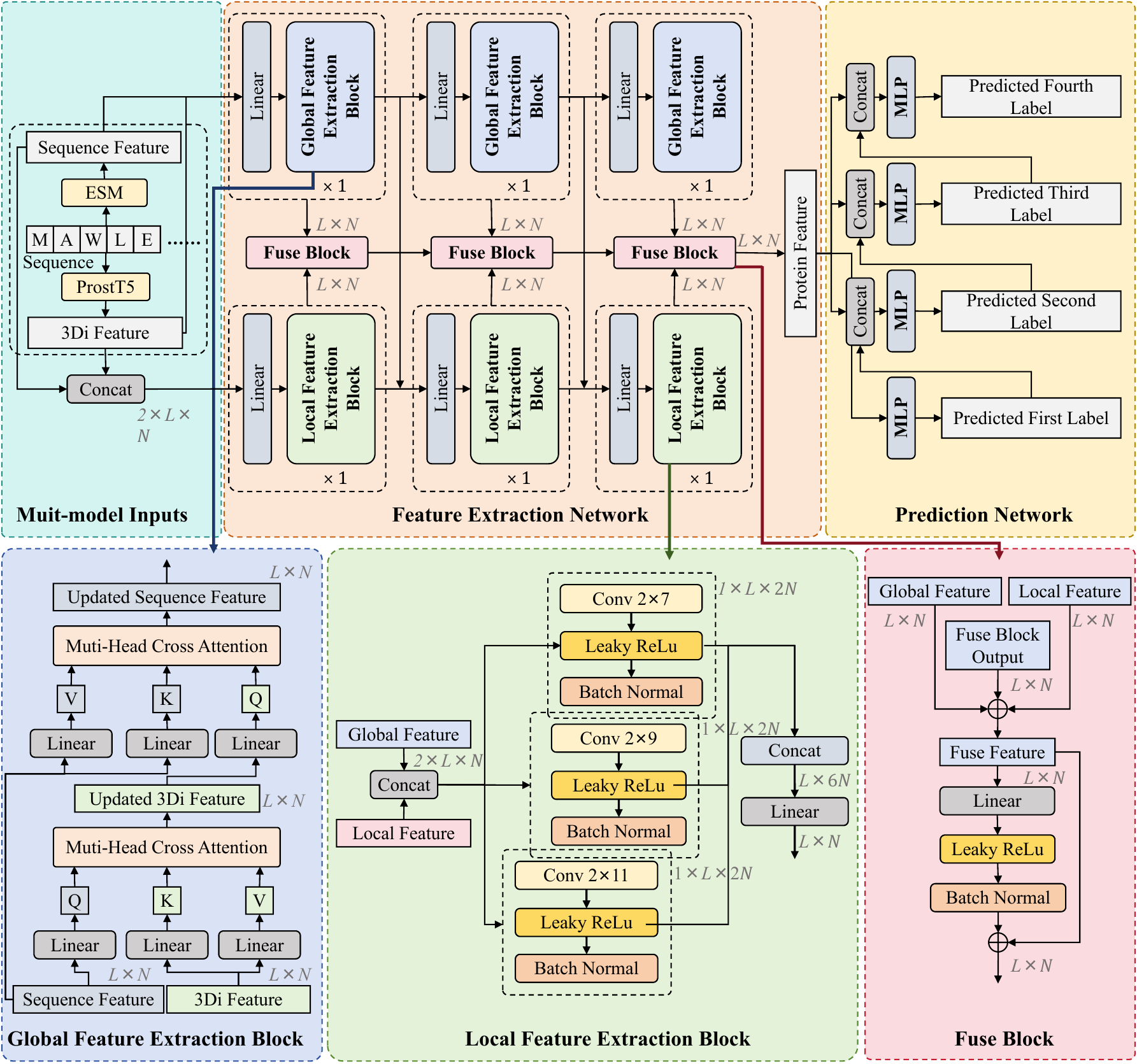}
       \vspace{-0.1in}
    \caption{Overview of MAPred. The inputs consist of the protein sequences and their corresponding 3Di tokens obtained through ProstT5. Within the Feature Extraction Network, we employ both a global feature extraction pathway and a local feature extraction pathway to capture the overall characteristics of the proteins and their specific functional sites, respectively. These features are then merged using a fuse block. In the Prediction Network, an autoregressive prediction architecture is utilized to predict the label for each digit in the EC number.}
    \label{fig:pipeline}
\end{figure}

We present the overall model framework in Figure \ref{fig:pipeline}. For the given protein sequences, we respectively use the pre-trained protein models ESM \cite{rives2021biological} and ProstT5 \cite{heinzinger2023prostt5} to obtain sequence features and corresponding 3Di features, and use them as inputs to the model. The architecture of MAPred consists of two networks: the Feature Extraction Network and the Prediction Network. The Feature Extraction Network comprises a global feature extraction pathway and a local feature extraction pathway. The global feature pathway is responsible for extracting the overall feature representations of the proteins, while the local feature pathway is designed to extract representations of functional sites. The Prediction Network operates on an autoregressive prediction scheme \cite{wehrmann2018hierarchical}. Instead of treating EC number prediction as a single multi-label classification problem, it predicts the digits of the EC number sequentially.
\subsection{Feature Extraction Network}
\subsubsection{Global Feature Extraction Block}
Our proposed Global Feature Extraction (GFE) pathway utilizes a strided sequence-to-3Di cross-attention mechanism to enhance the integration of protein sequence features with their corresponding structural features. As depicted in Figure \ref{fig:pipeline}, the global feature extraction pathway is constructed by stacking three GFE blocks, with each GFE block consisting of two cross-attention layers. In the first layer, the 3Di features are updated by incorporating features from the sequence. Conversely, in the second layer, the sequence features are enriched with characteristics derived from the updated 3Di representations. This bidirectional exchange of information allows for a comprehensive integration of both sequence and structural information.

Specifically, in the first cross-attention layer of $i$-th block, the sequence features $F_{global_{i}}^{seq} \in \mathbb{R}^{L \times N}$ serve as the query vectors, while the 3Di features $F_{global_{i}}^{3Di}\in \mathbb{R}^{L\times N}$ act as the key-value pairs. Utilizing scaled dot-product attention, as described in \cite{vaswani2017attention}, we compute the \textit{Multi-Head Attention} as:

\begin{gather} 
    {\rm MHA_{1}}(Q_{seq},K_{3Di},V_{3Di}) = [ head_{1}, head_{2}, \cdots, head_{h}] \\
    head_{i} = {\rm Att}(QW^{Q}_{i}, KW^{K}_{i}, VW^{V}_{i}) \\ 
    {\rm Att}(Q,K,V) = {\rm Softmax}(\frac{QK^{T}}{\sqrt{d_{k}}})V 
\end{gather} 

Here, $Q_{seq}\in \mathbb{R}^{L_{q} \times N}$, $K_{3Di}\in \mathbb{R}^{L_{k} \times N}$, and $V_{3Di}\in \mathbb{R}^{L_{v} \times N}$ represent the query, key, and value matrices, respectively, derived from $F_{global_{i}}^{seq}$ and $F_{global_{i}}^{3Di}$. $W^{Q}_{i}\in \mathbb{R}^{N \times N_{q}}$, $W^{K}_{i}\in \mathbb{R}^{N \times N_{k}}$, $W^{V}_{i}\in \mathbb{R}^{N \times N_{v}}$ and $N_{q} = N_{k} = N_{v} = N/h$ are learnable parameters. The $ [\cdot]$ means the concatenate operation. The attention mechanism allows the model to focus on the most relevant parts of the protein sequence when considering the 3Di structural information.

The output of the first cross-attention layer is a set of updated 3Di features $F_{global_{i}}^{\hat{3Di}}$, which now contains enriched information from the protein sequences. These updated features $F_{global_{i}}^{\hat{3Di}}$ are then used as input to the second layer, where the roles are reversed, the updated 3Di features become the query, while the original sequence features act as the key-value pairs:
\begin{equation}
    {\rm MHA_{2}}(Q_{\hat{3Di}},K_{seq},V_{seq}) = [ head_{1}, head_{2}, \cdots, head_{h}]
\end{equation}

This second layer further refines the sequence representations by aligning them with the structural features highlighted through the attention process. $F_{global_{i}}^{\hat{3Di}} \in \mathbb{R}^{L \times N} $ and $F_{global_{i}}^{\hat{seq}} \in \mathbb{R}^{L \times N} $ are the outputs of the GFE block. The sequential application of these two cross-attention layers within each GFE block enables MAPred to iteratively refine the feature representations, leading to a more nuanced and biologically meaningful fusion of sequence and structure.

\subsubsection{Local Feature Extraction Block}
Functional sites are conserved regions within proteins that serve specific functions and are crucial to both the structure and functionality of proteins \cite{bork1996protein}. To gain an in-depth understanding of the functional site characteristics within these proteins, similar to prior research \cite{li2020motifcnn,zeng2020protein}, we have employed a CNN-based approach in the Local Feature Extraction (LFE) pathway to construct contextual features for each amino acid. This method enables the model to account for the interactions between each amino acid and its neighbors, thereby enhancing the precision of functional site identification. As shown in Figure \ref{fig:pipeline}, the LFE pathway consists of three LFE blocks.

With the exception of the first block, the $i$-th block receives an input $F_{local_{i}}^{input} \in \mathbb{R}^{2 \times L \times N}$, which is a concatenation of the global sequence feature $F_{global_{i-1}}^{\hat{seq}} \in \mathbb{R}^{L \times N}$ and the local sequence features $F_{local{i-1}}^{output} \in \mathbb{R}^{L \times N}$. Each LFE block consists of three parallel convolutional networks, each containing convolution kernels of different sizes—7, 9, and 11. The output $F_{local_{i}}^{output}$ from each block is computed as the concatenation of the feature maps produced by these convolutional networks:

\begin{equation}
    F_{local_{i}}^{output} = {\rm MLP}([{\rm C}_{7}(F_{local_{i}}^{input}),{\rm C}_{9}(F_{local_{i}}^{input}),{\rm C}_{11}(F_{local_{i}}^{input})])
\label{equ:local}
\end{equation}

Here, ${\rm C}_{k}$ represents a convolution operation with a kernel size of $k$. This design provides an adaptive mechanism to extract both short-range and long-range dependencies within the protein sequence. The final output of the feature extraction network combines the local features with the global features, integrating detailed local functional sites with the overall sequence context to achieve a comprehensive understanding of proteins.

\subsection{Prediction Network}
After obtaining the integrated features $F_{fuse}=[F_{global_{3}}^{\hat{seq}}, F_{local_{3}}^{output}]$, considering that the EC number consists of four digits with strong sequential dependencies, we adopted an autoregressive prediction strategy to progressively predict each digit. Specifically, we first predict the first digit of the EC number, then use the predicted digit as an input feature for subsequent predictions, and so on, sequentially predicting the second, third, and fourth digits.

To implement this autoregressive prediction approach, we designed a multi-task learning framework comprising four MLPs, each responsible for predicting one digit. During the prediction process, the input features for ${\rm MLP}_{j}$ include both $F_{fuse}$ and the prediction results $\hat{y}_{j-1}$ from the previous ${\rm MLP}_{j-1}$. Formally, the autoregressive prediction process can be represented as:

\begin{equation}
\hat{y}_i =
\begin{cases}
{\rm MLP}_{i}(F_{fuse}; \theta_{i}),  & \text{if $i=1$} \\
{\rm MLP}_{i}(\left[\hat{y}_{i-1}, F_{fuse}\right]; \theta_{i}), & \text{otherwise}
\end{cases}
\label{equ:prediction}
\end{equation}

Here, $\theta_{j}$ represents the corresponding MLP parameters. Through this approach, each MLP can utilize the predicted results from the preceding MLP, thus better modeling the dependency relationships between labels.

\subsection{Model Training}

\begin{figure}
    \centering
    \includegraphics[width=0.95\linewidth]{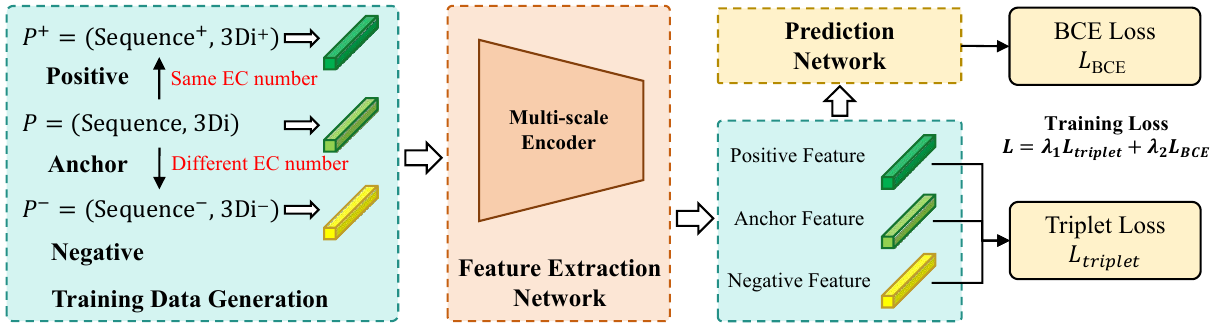}
       \vspace{-0.1in}
    \caption{The training stage of our model.}
    \label{fig:training}
\end{figure}

\subsubsection{Training Loss} A combined training loss is utilized which is constituted by 1) the triplet loss between samples of different classes, and 2) the BCE loss between the ground-truth and predicted EC number. Details are as follows. We use $L_{triplet}$ to denote the triplet loss between an anchor $a_i$, its positive sample $p_i$ and its negative sample $n_i$, which is mathematically defined as
\begin{equation}
L_{triplet} = \sum_{i=1}^{N} \max(0, d(a_i, p_i) - d(a_i, n_i) + margin)
\label{equ:contrastive}
\end{equation}
where $d(\cdot, \cdot)$ represents the distance, $margin$ is a predefined constant, and $N$ is the total number of triplets. In the meantime, we use $L_{BCE}$ to denote the label prediction loss/error between the predicted and the ground-truth EC number, which is mathematically defined as,
\begin{equation}
L_{BCE}(y_{i,j}, \hat{y}_{i,j}) = \sum_{j=1}^{4} \sum_{i=1}^{N} -\left[y_{i,j} \log(\hat{y}_{i,j}) + (1 - y_{i,j}) \log(1 - \hat{y}_{i,j})\right]
\label{equ:bce}
\end{equation}
where $\hat{y}_{i,j}$ refers to EC number predicted by the $j$-th MLP and $y_{i,j}$ refers to the actual EC number. The final combined loss could thus be formulated as
\begin{equation}
    L_{total} = \lambda_{1} \cdot L_{triplet} + \lambda_{2} \cdot L_{BCE}
\label{equ:loss}
\end{equation}
Here, $\lambda_{1}$ and $\lambda_{2}$ are weighting factors that take different values during different training phases, as specified in the following.

\subsubsection{Two-phase Training} We adopted a two-phase training scheme. In the first phase, we just train the feature extraction network using triplet learning, and the $\lambda_{2}$ in the loss function \ref{equ:loss} is set to 0. Once the feature extraction network converges, we start the second phase of training. In the second phase, we train the EC number prediction network, by setting the $\lambda_{1} = 0$ and $\lambda_{2} = 1.0$ in the loss function.

\section{Experiments}\label{experiments}

In this section, we first give a detailed description of the experimental protocol for this study. We then present comprehensive experiments to demonstrate: 1) the superior performance of our proposed framework for EC number prediction; 2) the effectiveness of the proposed components; and 3) MAPred can learn the functional regions of enzymes.

\subsection{Experimental Protocol}
\subsubsection{Datasets} We use the same Swiss-Prot dataset and data splitting as CLEAN \cite{yu2023enzyme}, encompassing 227,362 protein sequences that cover 5,242 EC numbers. To evaluate the performance of MAPred on real-world datasets, we employed two distinct datasets that are utilized by CLEAN during testing: New-392, Price-149 as well as a dataset we collected from Swiss-Prot named New-815. The New-392 dataset consists of 392 protein sequences, covering 177 different EC numbers. The Price dataset, as described by Price et al. \cite{price2018mutant}, is a collection of experimentally verified results, comprising 149 protein sequences that cover 56 different EC numbers. The New-815 dataset was collected using a method aligned with the data collection approach employed in the CLEAN, and consists of 815 protein sequences, covering 380 different EC numbers, containing data from Swiss-Prot released after April 2022.

\subsubsection{Implementation Details}\label{details} In the first training phase, we train the feature extraction network with a batch size of 40 for 1000 epochs. The learning rate is set to $5e - 4$ and remains constant during the training process. In the second training phase, we freeze the feature extraction network and train EC number prediction network with a batch size of 50000 for 150 epochs, the learning rate is $1e - 3$. We use Adam \cite{kingma2014adam} optimizer with $p_{1} = 0.9, p_{2} = 0.999$ and use the CosineAnnealingLR \cite{loshchilov2016sgdr} as the scheduler. Our approach is implemented with PyTorch \cite{paszke2019pytorch} framework, and all experiments are conducted on a machine with three NVIDIA RTX6000 ADA GPUs with AMD EPYC 7763 CPU and 256 G memory.

\subsection{EC Number Prediction}
We evaluate the accuracy of EC number prediction based on the commonly used Precision, Recall and F1 metrics, which are widely adopted in classification tasks. For benchmarking, we compare the performance of our proposed model with state-of-the-art classification models including: DeepEC \cite{deepec}, DeepECTF (short for DeepECtransformer) \cite{DeepECtransformer}, ProteInfer \cite{ProteInfer}, TFPC \cite{EnzBert}, CLEAN \cite{yu2023enzyme}.

\begin{table}
  \caption{Quantitative comparison of MAPred with the state-of-the-art EC number prediction methods. Prec denotes precision.}
  \label{tab:results}
  \centering
  \begin{tabular}{ccccccccccc}
    \toprule
    \multirow{2}{*} {\textbf{Method}} & \multicolumn{3}{c}{\textbf{New-392}} & \multicolumn{3}{c}{\textbf{Price}} & \multicolumn{3}{c}{\textbf{New-815}} \\
    \cmidrule(r){2-4}\cmidrule(r){5-7}\cmidrule(r){8-10}
    &\textbf{Prec}$\uparrow$&\textbf{Recall}$\uparrow$&\textbf{F1}$\uparrow$ & \textbf{Prec}$\uparrow$&\textbf{Recall}$\uparrow$&\textbf{F1}$\uparrow$& \textbf{Prec}$\uparrow$&\textbf{Recall}$\uparrow$&\textbf{F1}$\uparrow$ \\
    \midrule
    BlastP & {0.405} & {0.286} & {0.310}& {0.242} & {0.138}& {0.161} & {0.615} & {0.486}& {0.521}  \\
    \midrule
    CNN &{0.324} & {0.318} &{0.294}&  {0.235} & {0.184}& {0.197} &{0.502} & {0.523}& {0.495}\\
    RNN  &{0.167} & {0.163} & {0.148}& {0.102} & {0.078} &{0.067} &  {0.378} & {0.332}& {0.329}\\
    LSTM& {0.225} & {0.209} &{0.198} &{0.141} & {0.105} &{0.076}& {0.417} & {0.414}& {0.397} \\
    \midrule
    DeepEC  & {0.298} & {0.211} &{0.224}&  {0.164} & {0.079}&{0.096}& {0.471} & {0.273}& {0.309}\\
    DeepECTF  & {0.315} & {0.262} &{0.272} & {0.257} & {0.145}&{0.171}& {0.529} & {0.317}& {0.361} \\
    ProteInfer  & {0.358} & {0.245} &{0.262} & {0.184} & {0.086}&{0.106}& {0.504} & {0.309}& {0.346} \\
    TFPC  & {0.513} & {0.447} &{0.459} & {0.386} & {0.349}&{0.344}& {0.685} & {0.647}& {0.649} \\
    CLEAN  & {0.575} & {0.491} &{0.502} & {0.538} & {0.408}&{0.438}& {0.707} & {0.630}& {0.641} \\
    \midrule
    MAPred  & \textbf{0.651} & \textbf{0.632} & \textbf{0.610} & \textbf{0.554} & \textbf{0.487}& \textbf{0.493}& \textbf{0.721} & \textbf{0.683}& \textbf{0.680} \\
    \bottomrule
  \end{tabular}
\end{table}

\begin{figure}
    \centering
    \includegraphics[width=0.9\linewidth]{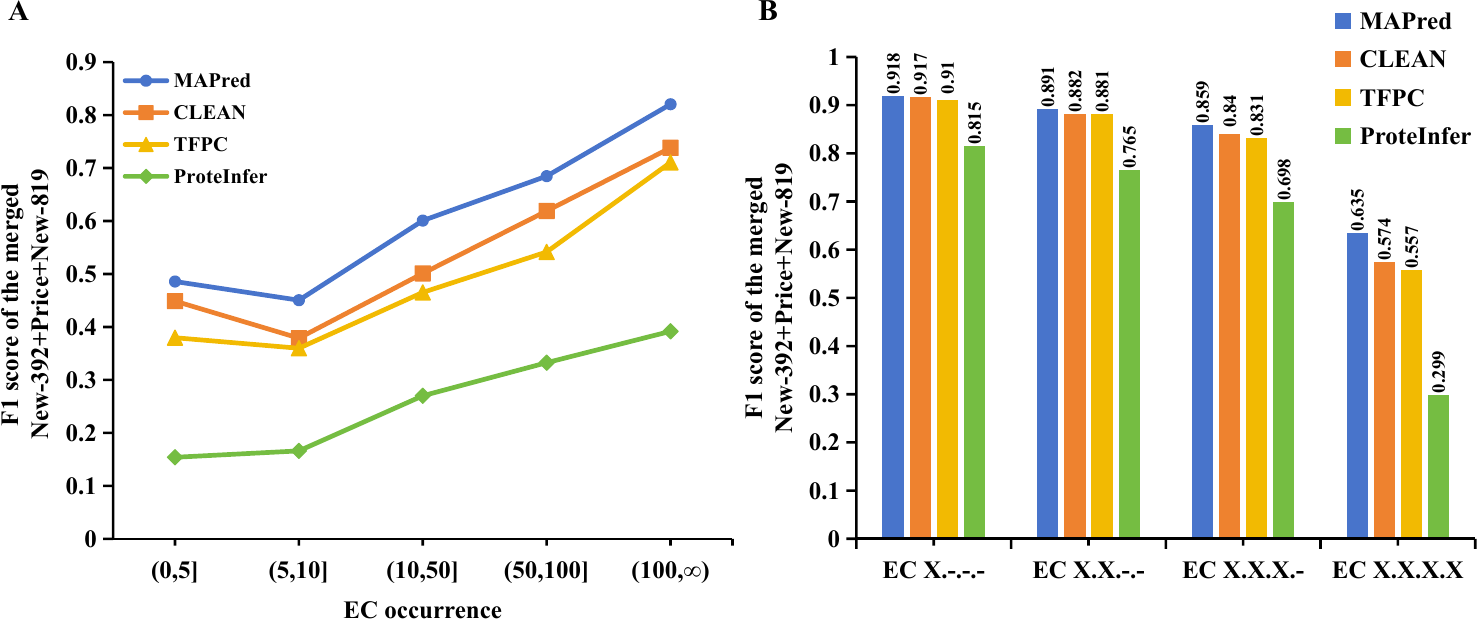}
       \vspace{-0.1in}
    \caption{Performance comparison of methods across different EC occurrence frequencies and hierarchical digit accuracy analysis. (A) Evaluation on the combined datasets binned by the number of times that the EC number appeared in training dataset. (B) Comparative analysis of the accuracy in predicting each digit of the EC number.}
    \label{fig:results}
\end{figure}

From the results shown in Table~\ref{tab:results}, we make the following important observations. Our method, MAPred, demonstrates a remarkable performance, outperforming existing approaches in all evaluated metrics. Specifically, on the New-392 dataset, MAPred achieves a Precision of 0.651, which is a significant improvement over the next best method, CLEAN, with a Precision of 0.575. This indicates that MAPred is highly accurate in predicting the correct EC numbers without a substantial number of false positives. In terms of Recall, MAPred also shows an impressive score of 0.632, which is higher than that of the CLEAN method, which has a Recall of 0.491, suggesting that MAPred is more effective in capturing the true positive instances. Furthermore, MAPred attains an F1 score of 0.610, again surpassing the CLEAN method's F1 score of 0.502. This result highlights the robustness of MAPred in achieving a good balance between Precision and Recall. Consistent performance gains are observed across the other datasets as well. For instance, on the Price dataset, MAPred's Precision, Recall, and F1 scores are 0.554, 0.487, and 0.493, respectively, which are superior to those of the CLEAN method. 

Furthermore, we present a detailed analysis of the results in Figure \ref{fig:results}. We use the F1 score to comprehensively evaluate the performance. Figure \ref{fig:results}(A) illustrates the methods' performance in predicting EC numbers with varying frequencies. It is evident that MAPred consistently outperforms the other methods across all ranges of EC occurrence. Notably, in the lower frequency intervals, such as (0, 5], MAPred maintains a leading position with an F1 score of about 0.486, indicating its superior ability to predict EC numbers in small sample sizes compared to other methods. In Figure \ref{fig:results}(B), the analysis focuses on the accuracy of predicting each of the four digits of the EC number. The results show that MAPred demonstrates high accuracy at each hierarchical level, starting with an F1 score of 0.918 for the first digit (EC X.$\_ . \_ . \_$) and maintaining a leading performance with an F1 score of approximately 0.635 for the final digit (EC X.X.X.X). Notably, MAPred, CLEAN, and TFPC exhibited similar F1 scores in predicting the initial three digits of the EC number. This similarity stems from the broader categorization of enzyme functions represented by these digits, which include a multitude of examples, thus simplifying the task for prediction models.

\subsection{Ablation Studies}

\begin{table}
  \caption{Ablation of MAPred modules. Prec denotes Precision. G, L, H denotes the global feature extraction pathway, the local feature extraction pathway, and the autoregressive label prediction model respectively.}
  \centering
  \begin{tabular*}{\textwidth}{@{\extracolsep{\fill}} ccccccccccc}
    \toprule
    \multirow{2}{*} {\textbf{Method}} & \multicolumn{3}{c}{\textbf{New-392}} & \multicolumn{3}{c}{\textbf{Price}} & \multicolumn{3}{c}{\textbf{New-815}} \\
    \cmidrule(r){2-4}\cmidrule(r){5-7}\cmidrule(r){8-10}
    &\textbf{Prec}$\uparrow$&\textbf{Recall}$\uparrow$&\textbf{F1}$\uparrow$ & \textbf{Prec}$\uparrow$&\textbf{Recall}$\uparrow$&\textbf{F1}$\uparrow$& \textbf{Prec}$\uparrow$&\textbf{Recall}$\uparrow$&\textbf{F1}$\uparrow$ \\
    \midrule
    MAPred & \textbf{0.651} & \textbf{0.632} & \textbf{0.610} & \textbf{0.554} & \textbf{0.487}& \textbf{0.493}& \textbf{0.721} & \textbf{0.683}& \textbf{0.680} \\
    \XSolidBrush G \checkmark L \checkmark H   & {0.354} & {0.447} &{0.303} & {0.316} & {0.283} & {0.282}& {0.455} & {0.414}& {0.378} \\
    \checkmark G \XSolidBrush L \checkmark H   & {0.583} & {0.573} &{0.548} & {0.536} & {0.441}&{0.463}& {0.706} & {0.651}& {0.652} \\
    \checkmark G \checkmark L \XSolidBrush H   & {0.611} & {0.604} &{0.558} & {0.550} & {0.467}&{0.481}& {0.719} & {0.681}& {0.674} \\
    \midrule
    Seq-only   & {0.630} & {0.517} &{0.538} & {0.542} & {0.467}&{0.476}& {0.716} & {0.630}& {0.649} \\
    3Di-only   & {0.575} & {0.491} &{0.505} & {0.540} & {0.408}&{0.439}& {0.679} & {0.515}& {0.545} \\
    \bottomrule
  \end{tabular*}
  \label{tab:ablation}
\end{table}
To verify the effectiveness of the proposed components, we conduct the following component contribution analysis experiment. As introduced above, MAPred mainly includes three components: 1) the global feature extraction pathway (denoted as \emph{G}); 2) the local feature extraction pathway (denoted as \emph{L}); and 3) the autoregressive label prediction model (denoted as \emph{H}). During the experiment, we disable one of these three components and re-train the remaining network parameters. Besides, to test the impact of multi-modality inputs on the results, we removed one part of the input, obtaining two results with only one type of input each, namely, seq-only (short for sequence-only) and 3Di-only. The study results are presented in the Table \ref{tab:ablation}. Note that we use tick/cross to indicate whether a certain component is enabled or disabled. 

From Table~\ref{tab:ablation}, we make the following observations: 1) Once the global feature extraction pathway (\emph{G}) is removed, the performance in predicting the EC number drops significantly. For example, on the New-392 dataset, the precision drops from 0.651 to 0.354. This substantial decline highlights the importance of the global feature extraction pathway, which captures the overall characteristics and long-range dependencies within the protein sequence.

2) The local feature extraction pathway (\emph{L}) also plays a significant role, as indicated by the performance decrease when it is disabled. However, the drop is less severe compared to the removal of the global pathway, suggesting that while local features contribute to the model's performance, the global context provides a more substantial impact on the overall prediction accuracy.

\begin{figure}[h]
    \centering
    \includegraphics[width=0.9\linewidth]{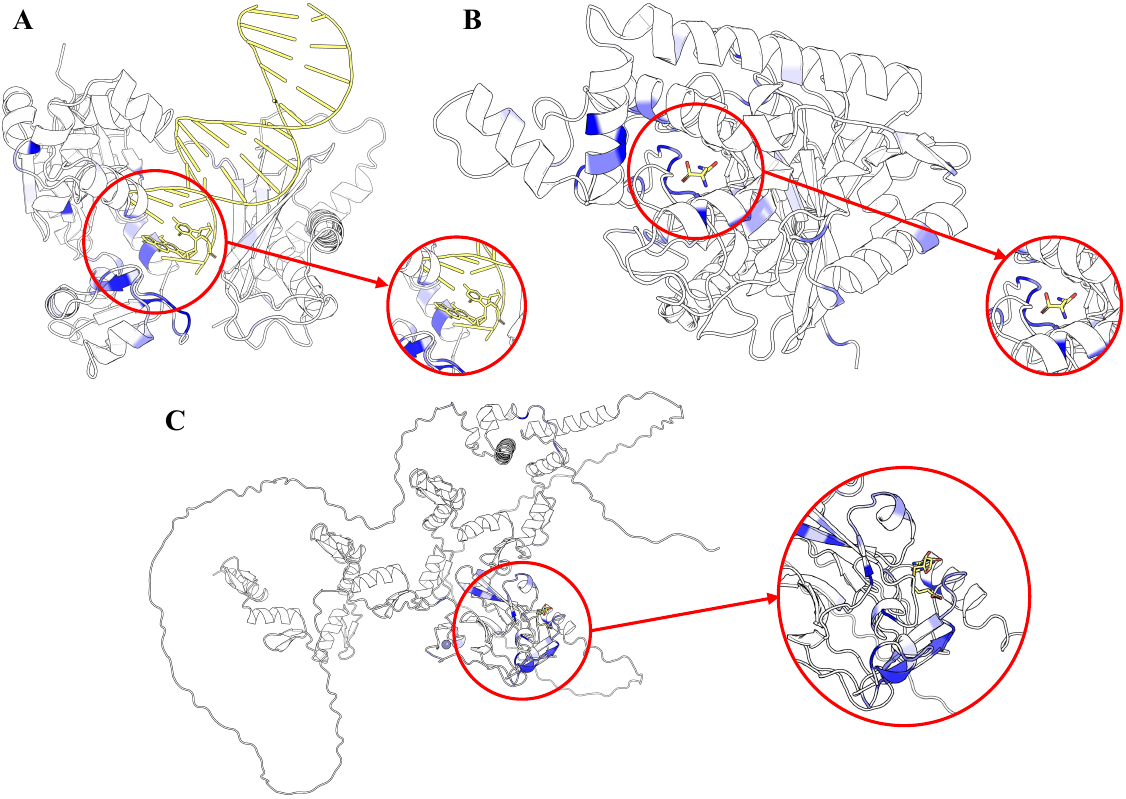}
       \vspace{-0.1in}
    \caption{Highlighted amino acid residues by the MAPred. The residues in blue indicate where the model pays more attention. (A) DNA polymerase IV dinB (UniProt ID: B8FBE8) with the catalytic center where polymerase reactions occur. (B) Glutamyl-tRNA amidotransferase gatA (UniProt ID: Q21RH9), with the reaction center and substrate glutamate in yellow. (C) Histone-lysine N-methyltransferase PRDM9 (UniProt ID: Q96EQ9), illustrating the reaction center where substrate lysine is modified.}
    \label{fig:vis}
\end{figure}

3) The autoregressive label prediction model (\emph{H}) shows its importance when considering the sequential dependencies among the four digits of the EC number. Disabling this component leads to a decrease in performance, emphasizing the value of treating the EC number as a hierarchical label rather than a flat label.

4) Multi-modality inputs also enhance the results, removing any modality leads to a decrease in performance metrics. Please note that MAPred performs better than CLEAN even with sequence-only input, demonstrating the model's superiority.

\subsection{MAPred Learns the Functional Regions of Enzymes}

MAPred can classify enzymes based on their EC numbers by utilizing the extraction of latent features from the amino acid sequences of enzymes. To assess whether MAPred has learned to identify the specific functional regions of enzymes, we analyzed the attention scores computed in the attention layer and visualized them on the three-dimensional structure of the enzyme, with the visualization depicted in Figure \ref{fig:vis}.

In Figure \ref{fig:vis}(A) and \ref{fig:vis}(B), the highlighted regions reveal that MAPred primarily concentrates on the catalytic sites where reactions occur. Specifically, Figure \ref{fig:vis}(A) shows attention to the polymerase reaction site, and Figure \ref{fig:vis}(B) highlights where the substrate binds and the reaction happens. Figure \ref{fig:vis}(C) illustrates the Histone-lysine N-methyltransferase PRD protein, characterized by a large part of disordered regions lacking specific functions. Here, the highlighted residues are predominantly and correctly located in the catalytic domain and near the substrate. These examples suggest that MAPred can identify potential reaction sites based on its attention scores, potentially influencing its predictions.

\section{Conclusion}
Our study presents MAPred, a novel approach to enzyme function prediction that integrates both sequence and structural data. The model's performance, as demonstrated through rigorous testing, surpasses existing methods, and achieves 0.610, 0.493, and 0.680 F1 scores on New-392, Price, and New-815 test sets, respectively. Our extensive ablation studies further validate the effectiveness of each component within MAPred, proving the effectiveness of our design. Future Work will focus on the excavation of key protein regions and the incorporation of more diverse datasets, expanding the prediction of labels to the Gene Ontology (GO) terms.



\bibliographystyle{unsrt}  

\end{document}